# Electric deflection of imidazole dimers and trimers in helium nanodroplets: Dipole moments, structure, and fragmentation

Benjamin S. Kamerin, John W. Niman, and Vitaly V. Kresin[*]

*Department of Physics and Astronomy, University of Southern California,*
*Los Angeles, CA 90089-0484, USA*

**Abstract**

Deuterated imidazole (IM) molecules, dimers and trimers formed in liquid helium nanodroplets are studied by the electrostatic beam deflection method. Monitoring the deflection profile of $(IM)D^+$ provides a direct way to establish that it is the primary product of the ionization-induced fragmentation both of $(IM)_2$ and $(IM)_3$. The magnitude of the deflection determines the electric dipole moments of the parent clusters: nearly 9 D for the dimer and 14.5 D for the trimer. These very large dipole values confirm theoretical predictions and derive from a polar chain bonding arrangement of the heterocyclic imidazole molecules.

[*] Corresponding author
Email *kresin@usc.edu*

## I. Introduction

Imidazole ("IM," $C_3H_4N_2$; see Fig. 1a) is an important, extensively studied constituent of proteins and biologically active compounds. The structural and electronic properties of imidazole complexes have also attracted a lot of interest because the molecule acts both as a proton donor (via its N–H constituent) and a proton acceptor (via the other nitrogen atom). It was recognized a long time ago[1] that this can enable IM to form linear structures in non-polar solvents. Furthermore, IM is a strongly polar[2] ($p=3.7$ D) and polarizable[3,4] ($\alpha=7.4$ Å$^3$) molecule and this, alongside the formation of the strong N–H$\cdots$N bond, has been predicted to endow IM oligomers with very large dipole moments[5-9] (see Figs. 1b,c). To the best of our knowledge, however, these dipole moments have not been directly measured up to now,[10] and this is one of the subjects of the present paper.

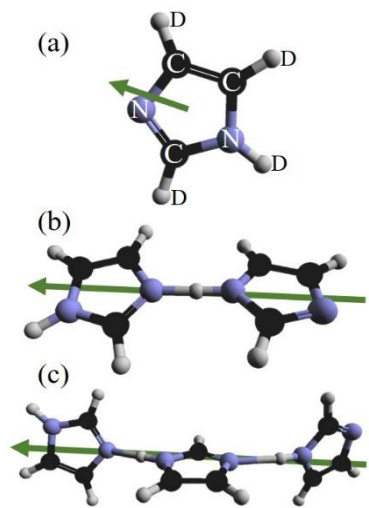

**Figure 1**. (a) Imidazole molecule, fully deuterated form. The arrow marks the molecule's electric dipole moment. (b,c) Linear dimer and trimer configurations, with calculated dipole moments (marked by arrows) of 9.1-9.6 D and 14.8 D, respectively.[7-9] The effect of deuteration on the ground-state dipole moments does not exceed a few percent.[2]

Experiments on the stability, spectroscopy, and photochemistry of gas-phase IM clusters, from dimers[6,7,9] to larger structures,[8,12] have been carried out to gain insight into the formation and structure of hydrogen-bonded IM complexes. An integral tool for studying size-dependent cluster properties is mass spectrometry, for which the particles need to be ionized. However, cluster ionization is typically accompanied by fragmentation and proton transfer, resulting in ions of uncertain parentage. The dominant products of electron-impact ionization of bare IM clusters are a series of $(IM)_nH^+$ peaks,[13] while clusters embedded in helium nanodroplets yield both $(IM)_nH^+$ and a weaker $(IM)_n^+$ series.[14] By applying the experimental approach described in this paper, we are able to identify the provenance of some of the small detected cluster ions.

The measurements described below are based on the technique of growing polar clusters inside superfluid helium nanodroplets by sequential pickup of molecules, and then deflecting the doped nanodroplet beam by a strong inhomogeneous electric field. This method has been demonstrated in our recent publications.[15,16] The amount of deflection associated with a detected ion determines its neutral parent's dipole moment and thereby the parent's identity.

**II. Experimental setup and mass spectra**

Publications 15,16 describe our setup and procedures in detail. A nanodroplet beam is formed by the expansion of pure helium gas through a cryogenic nozzle (15 K, 80 bar stagnation pressure). It traverses a cell filled with IM vapor where the molecular dopants are picked up. It is collimated by a 0.25 mm × 1.25 mm slit and passes between two 15 cm-long high voltage plates which create an electric field $E$ and a strong collinear field gradient directed perpendicular to the beam axis (82 kV/cm and 338 kV/cm², respectively). Then, after a 1.3 m field-free flight path, the beam enters an electron-impact ionizer (90 eV electron energy) through another narrow slit, and the resulting molecular ions are recorded by a quadrupole mass spectrometer synchronized with a beam chopper.

The ions are produced by charge exchange between the dopants and He$^+$ acceptors created within the nanodroplets by electron bombardment (see Refs. 15,17, and references therein). As mentioned above, the resulting energy release can be accompanied by substantial fragmentation of the dopant molecules. This process is governed solely by the dopant-acceptor interaction and therefore no difference in the relative ion intensity pattern was observed even at one-third of the above electron impact energy.

Deflections induced by the electric field on neutral doped nanodroplets are determined by setting the mass spectrometer to a particular ion mass and comparing its "field-on" and "field-off" spatial profiles. Both profiles are mapped out by translating the detector chamber, including its entrance slit, on a precision linear stage.

In order to better separate the peaks of IM and its fragments in the mass spectra, we used fully deuterated imidazole, IM$_D$ (CDN Isotopes, 98% purity). Since it has a high vapor pressure at room temperature, the powder could not be placed directly into the pick-up cell. Instead, IM resided in a heated glass vessel outside the chamber and its vapor was fed into the cell through a heated narrow tube and heated needle valves. With careful heating we were able to achieve stable signals for the 4-6 hour duration of a full deflection measurement cycle. Molecules are picked up by helium nanodroplets via successive collisions in a Poisson process.[18] Deflections were measured for two vessel temperatures, 30ºC and 50ºC, corresponding to what will be referred to as "lower" and "higher" droplet doping regimes. As described below, in the former case we identify the majority of resulting dopant formations to be IM dimers, and in the latter case trimers.

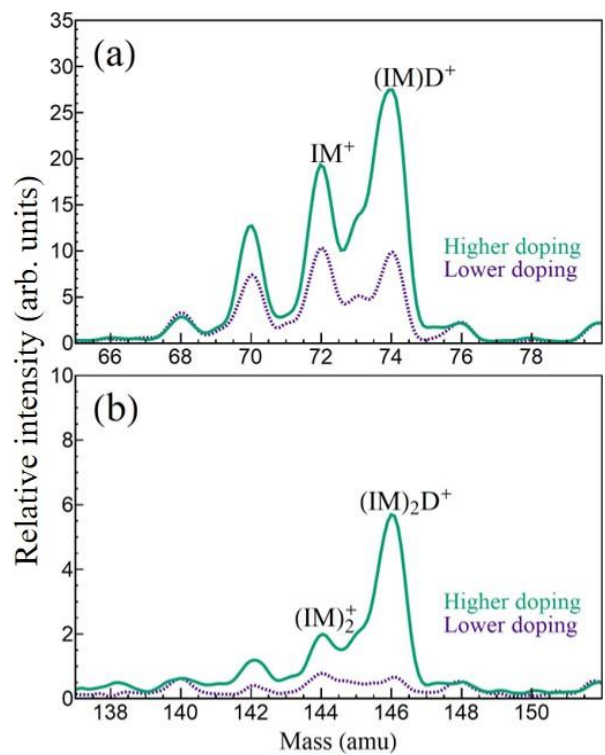

**Figure 2**. Mass spectra produced by electron impact ionization of helium nanodroplets doped with deuterated imidazole, $IM_D$. The top (a) and bottom (b) panels show the monomer and dimer regions, respectively. The lower (dashed) lines correspond to the same low-doping regime and the upper (solid) lines to the same higher-doping regime, see the text, and the vertical scale of the panels matches the relative magnitudes of all the peaks. The additional signal visible between the labeled ion peaks is believed to derive from partially undeuterated $IM_H$ present in the original powder and from possible H/D exchange occurring within the vapor supply system.

Fig. 2 shows the mass peaks in the monomer and dimer regions at two different IM doping levels. The peak at 72 amu corresponds[19] to the molecular ion $(C_3D_4N_2)^+$ and the one at 74 amu to $(C_3D_4N_2)D^+$. We see that the latter grows far more rapidly with increased molecular vapor pressure, indicating that it derives from progressively larger complexes forming inside the droplet.

In the lower-doped regime the $(IM)_3^+$ peak was virtually absent in the mass spectrum. Previously published mass spectra[14] suggest that ~5%-10% of imidazole cluster ionization products in helium nanodroplets are detected as $(IM)_n^+$, therefore the absence of any trimer signal implies that in this case the $(IM)D^+$ signal originates primarily from dimers. Furthermore, the $(IM)_2^+$ peak is much weaker than $(IM)D^+$, therefore the latter is the dominant dimer ionization channel.

In the higher-doped regime both $(IM)_2D^+$ and $(IM)D^+$ grow substantially, implying that the droplets now contain many trimers (and possibly higher order oligomers) which therefore contribute the main portion of the $(IM)D^+$ signal.

As remarked in the Introduction, the deflection method makes it possible to support such considerations in a more quantitative manner. The data discussed below confirm that the $IM^+$ signal derives primarily from monomers, while dimers and trimers indeed undergo extensive fragmentation and are detected in the $(IM)D^+$ channel.

There is also a peak at 70 amu corresponding to deuterium loss, $(C_3D_3N_2)^+$. Interestingly, this peak does not appear nearly as prominently in the electron-impact mass spectrum of pure gas-phase imidazole.[20] However, it is known that ionization branching ratios within nanodroplets are not always identical to those of free molecules.[21,22] Both the pressure dependence of this peak's intensity and its deflection profile suggested that it is primarily a product of the ionization of IM monomers.

**III. Electric deflections**

Polar molecules and complexes located within the extremely cold ($T$=0.37 K) superfluid helium nanodroplets become almost completely oriented along the electric field of the electrodes. This effect was established by landmark spectroscopic measurements of pendular dopant states in nanodroplets.[23] Our experiments take advantage of the fact that the oriented dipoles experience such a strong force from the field's gradient that the entire beam of heavy (~$10^4$ He atoms) doped nanodroplets deflects by a substantial angle. These experiments have confirmed that the dopants are highly oriented by demonstrating that the amount of deflection is linearly proportional to the deflection voltage[15,24] (otherwise it would be proportional to the voltage squared[25,26]).

As described above, deflection measurements require tight beam collimation, hence we need to select sufficiently intense peaks in the mass spectrum. In the present case this is fulfilled for the $IM^+$ and $(IM)D^+$ peaks. Fig. 3 shows their deflection profiles under the lower and higher doping conditions.

One can immediately see from the figure that the deflection of the deuterated $(IM)D^+$ ion (corresponding to the 74 amu peak in Fig. 2a) significantly exceeds that of the molecular ion (72 amu) and, furthermore, that this deflection is stronger at the higher pressure. This matches the conclusion above that this peak in the mass spectrum derives from larger and more polar intact molecular complexes.

The magnitudes of these complexes' dipole moments can be determined from the amount of deflection shown by the profiles in Fig. 3.

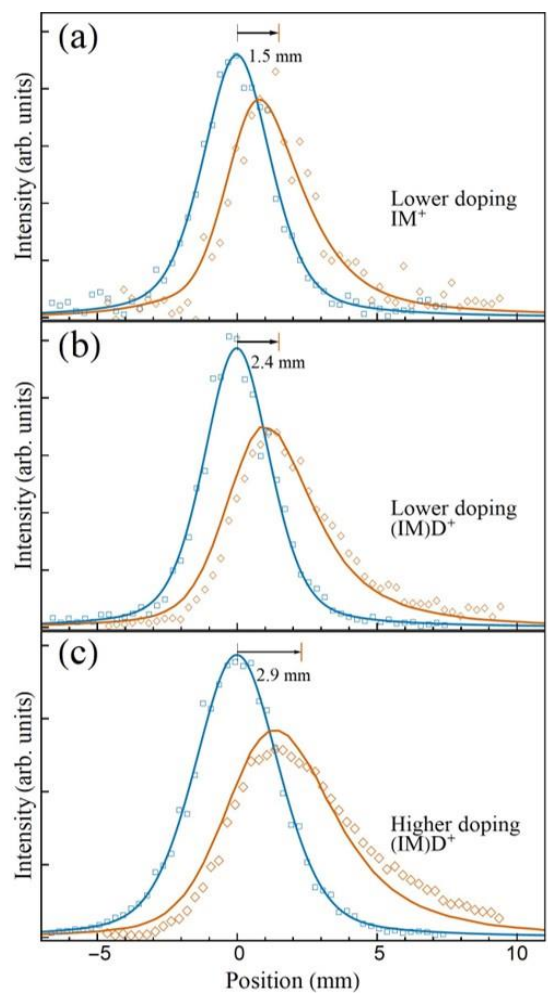

**Figure 3.** Electrostatic deflection of $(IM)_n$-doped helium nanodroplet beams. Blue: electric field off, orange: electric field on. The symbols are experimental data, the lines are fits of the deflection process, as described in the text. (a) $IM^+$ signal; (b),(c) $(IM)D^+$ signal in the lower and higher doping regimes, respectively. The arrows denote the shift of the profile centroids.

In analyzing these profiles one must keep in mind that the nozzle source produces a log-normal distribution of nanodroplet sizes.[18] To calibrate this distribution we employ the 72 amu ion peak profiles in Fig. 3a, which is assigned to the IM monomer.[19] Its known dipole moment and rotational constants[2] are used to compute its orientation in the electric field. This is done by diagonalizing the rigid rotor Stark Hamiltonian matrix which incorporates the rotational states occupied at the 0.37 K temperature of the helium droplet matrix. It is known that that for many molecules their rotational constants are reduced[18,23] by a factor of 2.5–3 due to coupling to the superfluid.[27] We applied this factor in our analysis; the end result for the orientation cosine, $\langle\cos\theta\rangle=0.90$, was insensitive to its precise value. The molecular polarizability term, $\alpha E$, can be neglected because it comprises less than 0.1% of the permanent dipole moment.

Next, a fit to the deflection profile, using a Monte Carlo simulation which incorporates the pick-up, evaporation, deflection, and ionization steps,[15] yields the nanodroplet size distribution parameters. The average size was $\bar{N}\approx 1.2\times 10^4$ and the width was $\Delta N\approx 1.0\times 10^4$. The ratio $\Delta N/\bar{N}$ is similar to the values observed in other experiments.[28] Using the He$_N$ diameter[18] $D\approx 4.4 N^{1/3}$ Å and the bond lengths of the IM molecule and dimer,[29] the average nanodroplet can be visualized as being roughly 25 monomers or 10 dimers across.

This information now can be used with the same simulation to deduce the electric dipole moments corresponding to the deflection profiles in Figs. 3b and 3c, i.e., the profiles measured with the mass spectrometer set to detect the (IM)D$^+$ ion. Since the corresponding electric dipole values are large, the orientation cosines used in these fits can be accurately represented by the Langevin-Debye function. The rigid chain model is applicable in view of the extremely low temperature: even the lowest-energy vibrational frequencies of the IM dimer (monomer rocking and twisting[29]) lie above 10 cm$^{-1}$, i.e., forty times above the thermal energy. This represents an interesting contrast with hydrogen-bonded complexes at higher temperatures, where flexible vibrational motion can lead to a sizeable electric dipole moment.[30]

The results are 8.8 D and 14.5 D for the "lower" and "higher" doping regimes, respectively, with an estimated error of ±0.6 D. Within the accuracy of the measurement, these values are persuasively in agreement with the predicted dipole moments of the imidazole dimer and trimer, listed in the caption of Fig. 1.

Higher-order oligomers are also present in the beam under the higher-doping conditions, as evidenced by the $(IM)_3D^+$ peak in the corresponding mass spectrum, see the Supplementary Material. However, the fact that this peak is very much weaker than $(IM)_2D^+$ and $(IM)D^+$ suggests that the admixture of larger clusters is small and does not strongly "contaminate" the deflection measured on the $(IM)D^+$ channel. The consistency between experimental and theoretical values supports this conclusion.

Therefore, two important deductions can be made:

First, the doped nanodroplet deflection method, in a direct and quantitative way, confirms the prediction of a remarkably strong dipole moments of these complexes, deriving from their aligned structure and facilitated by the formation of a strong hydrogen bond and mutual polarization. There is no evidence for the formation of cyclical $IM_3$ structures[31] within the nanodroplets: they would have manifested themselves as a secondary peak near the zero position in the deflection profile.

Second, the method unambiguously corroborates the fact that the $(IM)D^+$ ion is the dominant product of the ionization of both $IM_2$ and $IM_3$. This ability to ascertain fragment parentage is a useful and novel supplement to mass spectrometry.

## IV. Conclusions

In summary, we have applied the method of electrostatic beam deflection to helium nanodroplets doped with deuterated imidazole molecules. The deflection patterns provide express evidence that imidazole dimers and trimers extensively fragment upon ionization, with the dominant channel being the formation of the protonated imidazole ion. This fragmentation tracing technique can be extended to larger systems, by using precise vapor pressure control in the pick-up cell in order to generate a sequence of incrementally larger oligomers within the nanodroplets.

Absolute values of the electric dipole moments of the dimer and the trimer, not previously determined experimentally, also were determined from the measurement, and are consistent with theoretical predictions. These high dipoles derive from imidazole molecules arranging themselves into highly polar linear chains facilitated by a strong proton bond between the nitrogen atoms in adjacent rings. The same bond is responsible for the dominance of proton transfer upon excitation, as observed here.

Since the electric dipole moment is sensitive to the geometric structure of a molecule and its charge density distribution, deflection measurements can serve as a valuable complement to spectroscopy and mass spectrometry of molecular complexes, including oligopeptides and clusters of heterocyclic compounds. This has been illustrated, for example, in experiments on gas-phase peptides.[25] The use of helium nanodroplet isolation extends this approach by enabling the formation of a variety of complexes using successive pickup steps, and by simplifying the deflection analysis thanks to freezing out the vibrational degrees of freedom of nonrigid systems.

## SUPPLEMENTARY MATERIAL

The supplementary material contains an extended mass spectrum showing the range from the monomer to the trimer of imidazole under two nanodroplet doping conditions.

## DATA AVAILABILITY

The data that support the findings of this study are available within the article and from the corresponding author upon reasonable request.

## ACKNOWLEDGMENTS

B.S.K. and J.W.N. contributed equally to this work. We would like to thank M. Fárník, L. Kranabetter and S. W. Niman for productive discussions. This research was supported by the U.S. National Science Foundation under Grant No. CHE-1664601.